\pdfoutput=1
\documentclass[11pt,twoside,a4paper,cmspaper,final,collab]{cms-tdr}
\def\svnVersion{830f182}\def\svnDate{2025/03/17}\def\cmsCernNoTag{CERN-EP-2024-180}\def\cmsCernDate{\today}\def\cmsMessage{Published in Physical Review D as \href{http://dx.doi.org/10.1103/PhysRevD.111.L051102}{\doi{10.1103/PhysRevD.111.L051102}.}}
\begin{document}\cmsNoteHeader{BPH-22-012}

\newcommand{\PBcJpsimu}{\ensuremath{\PBpc\to\PJGy\,\PGmp\PGnGm}\xspace}
\newcommand{\PBcJpsitau}{\ensuremath{\PBpc\to\PJGy\,\PGtp\PGnGt}\xspace}
\newcommand{\rjpsi}{\ensuremath{R(\PJGy)}\xspace}
\newcommand{\Hb}{\ensuremath{\mathrm{H}_{\PQb}}\xspace}
\newcommand{\rjpsilonginline}{\ensuremath{R(\PJGy) = \mathcal{B}(\PBpc\to\PJGy\PGtp\PGnGt) / \allowbreak \mathcal{B}(\PBpc\to\nobreak\PJGy\PGmp\PGnGm)}\xspace}
\newcommand{\mMuMu}{\ensuremath{m_{\PGm\PGm}}\xspace}

\newlength\cmsTabSkip\setlength{\cmsTabSkip}{1ex}
\ifthenelse{\boolean{cms@external}}{\providecommand{\prlWidth}{0.95\columnwidth}}{\providecommand{\prlWidth}{0.6\textwidth}}
\ifthenelse{\boolean{cms@external}}{\providecommand{\cmsLeft}{upper\xspace}}{\providecommand{\cmsLeft}{left\xspace}}
\ifthenelse{\boolean{cms@external}}{\providecommand{\cmsRight}{lower\xspace}}{\providecommand{\cmsRight}{right\xspace}}
\ifthenelse{\boolean{cms@external}}{\providecommand{\NA}{\ensuremath{\cdots}\xspace}}{\providecommand{\NA}{\ensuremath{\text{---}}\xspace}}

\cmsNoteHeader{BPH-22-012}
\title{Test of lepton flavor universality in semileptonic \texorpdfstring{$\PBpc$}{Bc} meson decays in proton-proton collisions at \texorpdfstring{$\sqrt{s}= 13\TeV$}{sqrt(s) = 13 TeV}}

\date{\today}

\abstract{
\tolerance=2400
A measurement of the ratio of branching fractions $R(\PJGy) = \mathcal{B}(\PBpc\to\PJGy\PGtp\PGnGt) /  \mathcal{B}(\PBpc\to\PJGy\PGmp\PGnGm)$ in the $\PJGy\to\PGmp\PGmm$, $\PGtp\to\PGmp\PGnGm\PAGnGt$ decay channel is presented. This measurement uses a sample of proton-proton collision data collected at a center-of-mass energy of 13\TeV by the CMS experiment in 2018, corresponding to an integrated luminosity of 59.7\fbinv. The measured ratio, $R(\PJGy) = 0.17^{+ 0.18}_{- 0.17}\stat^{+ 0.21}_{- 0.22}\syst^{+ 0.19}_{- 0.18}\thy = 0.17 \pm 0.33$, agrees with the value of $0.2582 \pm 0.0038$ predicted by the standard model, which assumes lepton flavor universality.  By testing lepton flavor universality, this measurement is a probe of new physics using \PBpc mesons, which are currently only produced at the LHC.}

\hypersetup{
pdfauthor={CMS Collaboration},
pdftitle={Test of lepton flavor universality violation in semileptonic Bc meson decays at CMS},
pdfsubject={CMS},
pdfkeywords={CMS, lepton flavor violation, R(J/Psi), BPH}}

\maketitle 

In the standard model (SM), the three lepton families have the same couplings for electroweak interactions, an accidental symmetry known as lepton flavor universality (LFU). Differences in decay rates between processes that differ solely by lepton flavor thus originate only from the different lepton masses.

Several beyond the SM (BSM) models that contain additional particles and nontrivial flavor interactions, such as an extended Higgs sector~\cite{Tanaka:1994ay, Crivellin:2012ye}, leptoquarks~\cite{Freytsis:2015qca}, or an extended gauge sector~\cite{Crivellin:2015lwa}, predict LFU violation. To date, no direct evidence for the existence of BSM particles has been found, and constraints are set on these models, \eg, in Refs.~\cite{ATLAS:2021jyv,CMS:2021far, CMS:2022fsw}. However, even if they are too heavy to be produced at existing colliders, BSM particles could still contribute via virtual interactions and alter the decay rates predicted by the SM.

Lepton flavor universality has been confirmed in leptonic \PZ and \PW boson decays down to the per-mille level~\cite{UA1:1988rck, ALEPH:2005ab, ATLAS:2016nqi, ATLAS:2020xea, CMS:2022mhs} and, in recent years, has been extensively tested in semileptonic $\PQb$ hadron (\Hb) decays through the measurement of ratios of branching fractions. In particular, the BaBar~\cite{BaBar:2013mob}, Belle~\cite{Belle:2015qfa, Belle:2016ure}, and LHCb~\cite{LHCb:2015gmp, LHCb:2017smo, LHCb:2023zxo} Collaborations investigated the $\ensuremath{R(\PDst)} = \mathcal{B}(\PBz\to\PDstm\PGtp\PGnGt) / \mathcal{B}(\PBz\to \PDstm\PGmp\PGnGm)$ ratio, observing a combined value of $R(\PDst) = 0.295 \pm 0.014$, $3.2$ standard deviations above the SM expected value of $0.254 \pm 0.005$~\cite{HFLAV:2022pwe}. In this Letter, charge conjugate states are implied.

The same, possibly nonuniversal, lepton couplings in $\PQb \to \PQc \Pl \PGnl$ transitions can also be experimentally tested in semitauonic decays from \Hb hadrons other than \PBz.
In this Letter, we present the measurement of the ratio 
\begin{equation*}
\rjpsi = \frac{\mathcal{B}(\PBcJpsitau)}{\mathcal{B}(\PBcJpsimu)}.
\end{equation*}  
A recent calculation predicts a value of $\rjpsi = 0.2582 \pm 0.0038$~\cite{Harrison:2020nrv}, which is consistent with earlier estimates~\cite{Anisimov:1998uk, Ivanov:2006ni, Hernandez:2006gt, Azizi:2019aaf}.
Since \PBpc mesons cannot be produced at the existing B factories, this ratio has not been extensively explored. The only measurement to date, by the LHCb Collaboration~\cite{LHCb:2017vlu}, reports a two standard deviation excess over the SM prediction. 
The present measurement, which is also provided in a HEPData record~\cite{hepdata}, uses a sample of proton-proton collision data collected at a center-of-mass energy of 13\TeV by the CMS experiment in 2018, corresponding to an integrated luminosity of 59.7\fbinv~\cite{CMS-PAS-LUM-18-002}.

The CMS apparatus~\cite{CMS:2008xjf} is a multipurpose, nearly hermetic detector, designed to trigger on~\cite{CMS:2020cmk,CMS:2016ngn} and identify electrons, muons, photons, and (charged and neutral) hadrons~\cite{CMS:2020uim,CMS:2018rym,CMS:2014pgm}. A global ``particle-flow'' algorithm~\cite{CMS:2017yfk} aims to reconstruct all individual particles in an event, combining information provided by the all-silicon inner tracker and by the crystal electromagnetic and brass-scintillator hadron calorimeters, operating inside a 3.8\unit{T} superconducting solenoid, with data from the gas-ionization muon detectors embedded in the flux-return yoke outside the solenoid. The reconstructed particles are used to build \PGt leptons, jets, and missing transverse momentum, defined as the negative vector sum of the transverse momenta of all the reconstructed candidates in an event~\cite{CMS:2018jrd,CMS:2022prd,CMS:2016lmd,CMS:2019ctu}.

{\tolerance=600 The analysis reconstructs both the \PBcJpsitau and \PBcJpsimu signals in the $\PJGy\to\PGmp\PGmm$ and $\PGtp\to\PGmp\PGnGm\PAGnGt$ decay chains, thus resulting in the identical visible products $(\PGmp \PGmm)\PGmp$ for both processes, where the muon not originating from the \PJGy meson decay is referred to as the ``third'' muon. The signal final states differ in the number of neutrinos and the muon topology, owing to the tau lepton lifetime. Several background contributions are considered: processes where the \PJGy meson and the third muon come from either the same or different \PQb hadrons (including a \PBpc meson, excluding the signal modes), events where a hadron is misidentified as the third muon, and three-muon events with two unrelated opposite-charge muons with invariant mass $\mMuMu$ close to that of the \PJGy meson. Backgrounds are estimated from data control regions (CRs) or, when not possible, from simulations complemented with inputs from data. The expected signal and background contributions are fitted to data to extract \rjpsi through a maximum likelihood binned fit.\par}

Events are retained if they satisfy triggers requiring three muons with $\pt> 5$, 3.5 and 2\GeV, all within pseudorapidity $\abs{\eta} < 2.5$, the invariant mass of at least one opposite-charge muon pair to lie between 2.95--3.25\GeV, consistent with the world-average \PJGy meson mass $m_{\PJGy}=3.097\GeV$~\cite{PDG2022}, and the corresponding \textit{p}-value of the dimuon vertex fit greater than 0.5\%. 

In the offline data analysis, further requirements are applied to events satisfying the trigger selection.
The \PJGy candidates are required to have a vertex fit \textit{p}-value above 1\% and a reconstructed mass within 50\MeV of $m_{\PJGy}$ if all muons have $\abs{\eta}<1$, 100\MeV if they have $\abs{\eta}>1$, or 70\MeV otherwise.
In each event, the three muon candidates must have $\pt>4\GeV$ and $\abs{\eta}<2.4$, and at least one must have $\pt>6\GeV$. The two muons forming the \PJGy candidate must satisfy the ``medium'' working point of the standard CMS muon identification algorithm~\cite{CMS:2018rym}.

{\tolerance=300 For each 3\PGm candidate, its corresponding primary vertex (PV) is defined as the one with the shortest longitudinal distance with respect to the \PJGy meson momentum direction back-propagated onto the beam line.
The transverse impact parameter (defined as the distance of closest approach of the particle's trajectory to the beam line) of each muon track is required to be smaller than $0.5\unit{mm}$ to ensure its compatibility with coming from a decay of a \PBpc meson.
For any pair of muons, the absolute difference between their longitudinal impact parameters must be smaller than $2\unit{mm}$ to further reduce accidental combinations of muons coming from different PVs.
Finally, the three muons are required to have a vertex fit \textit{p}-value above 0.01\%, with the two muons forming the \PJGy candidate constrained to the \PJGy mass.\par}

Kinematical variables sensitive to the large mass difference between tau leptons and muons and to the presence of two additional neutrinos in the $\PGtp\to\PGmp\PGnGm\PAGnGt$ decay allow disentangling the two signals and help to identify and reduce backgrounds.
The determination of such variables benefits from the knowledge of the original \PBpc candidate four-momentum in the laboratory frame ($p_{\PBpc}$), which can only be inferred indirectly, due to undetected neutrinos.
The \PBpc four-momentum is estimated from the 3\PGm candidate four-momentum $p_{3\PGm}^{\text{vis}}$ and mass $m_{3\PGm}^{\text{vis}}$ as $p_{\PBpc} = (m_{\PBpc}/m_{3\PGm}^{\text{vis}}) p_{3\PGm}^{\text{vis}}$, where $m_{\PBpc}$ is the \PBpc meson mass~\cite{PDG2022}. The direction of $p_{\PBpc}$ is therefore assumed to be aligned with that of $p_{3\PGm}^{\text{vis}}$.
A variable that efficiently discriminates between the final states of the two signal decays is the squared four-momentum transfer to the lepton system $q^2=(p_{\PBpc} - p_{\PJGy})^2$, where $p_{\PJGy}$ is the four-momentum of the \PJGy candidate.
In addition, two topological observables are also used: $\mathrm{IP3D}/\sigma_{\mathrm{IP3D}}$, where the 3D impact parameter ($\mathrm{IP3D}$) is the shortest distance from the third muon track to the \PJGy vertex with a sign depending on whether the trajectory of the third muon reaches the minimum distance before or after the \PJGy vertex along the \PJGy candidate momentum direction, and $\sigma_{\mathrm{IP3D}}$ is its uncertainty; and $L_{xy}/\sigma_{L_{xy}}$, where $L_{xy}$ is the distance between the \PJGy meson decay vertex and beam line in the transverse plane, and $\sigma_{L_{xy}}$ is its uncertainty.

Simulated samples are used to model all contributions yielding a \PJGy meson in association with a genuine \Pgmp from either the same or different decay chains, as opposed to a misidentified muon (misID), defined as a muon from \PKp or \PGpp decays, photon conversions, or misreconstructed hadrons. Processes with misID or where the \PJGy candidate arises from accidental muon pairs are estimated from data, as discussed later.
Two separate Monte Carlo (MC) samples describe decays initiated by a \PBpc meson or another \Hb. 

In the first sample, \PBpc mesons are generated with the dedicated \textsc{bcvegpy}~2.2b~\cite{Bcveg} software. Decays of interest are forced using \EVTGEN~1.6.0~\cite{Lange:2001uf} and the event hadronization is handled by \PYTHIA~8.240~\cite{SJOSTRAND2015159}. 
The set of forced decays includes the signal channels, \PBcJpsimu and \PBcJpsitau (all \PGtp decays are allowed and $\mathcal{B}(\PGtp\to\PGmp\PGnGm\PAGnGt) = 17.41\%$), $\ccbar$ states that decay into the \PJGy meson, and other $\PBpc\to\PJGy\PD^{(*)}_{(\mathrm{s})} \PX$ decays, where \PX indicates up to two extra hadrons.
The only feed-down processes with sizable contributions after the final selection are $\PBpc\to\PGyP{2S}\PGmp\PGnGm$, $\PBpc\to\PGccDoP{1P}\PGmp\PGnGm$, and $\PBpc\to\PGccDtP{1P}\PGmp\PGnGm$. While $\PBpc\to\PGyP{2S}\PGtp\PGnGt$ is also considered in the analysis, its final yield is negligible.
The branching fractions of the various forced decays are expressed relative to $\mathcal{B}(\PBcJpsimu)$ and fixed to their measured values~\cite{PDG2022,LHCb:2019tea,LHCb:2013kwl,ATLAS:2015jep,LHCb:2016djy} or, lacking thereof, to theoretical predictions or upper limits.
The properties of the interaction between the \PBpc meson constituent quarks are described by form factors, which govern the decay kinematics and therefore impact the $q^2$ observable. In this sample, for the \PBcJpsitau and \PBcJpsimu processes, the form factors predicted in Ref.~\cite{KISELEV2000473}
are used. A generator-level reweighting procedure based on the \textsc{Hammer} package~\cite{Bernlochner:2020tfi} is employed to update the form factors to more recent predictions based on the lattice quantum chromodynamics for semileptonic $\PQb\to\PQc\Pell\PGn$ transitions~\cite{PhysRevD.100.094503}.

In the second sample, $\Hb\to\PJGy(\to\PGmp\PGmm) + \PGmp$ processes are simulated. The third muon can either come from the same  \Hb as the \PJGy meson or from an unrelated source. The \PYTHIA generator is used to produce and hadronize $\PQb$ quark pairs into \Hb, then \EVTGEN is used to enforce decays of interest.
To improve the generation efficiency, \Hb are forced to decay in modes that contain a \PJGy meson (including feed-downs). The list of \Hb that are subject to forced decays includes \PBp, \PBz, \PBzs, \PGLzb, \PGXmb, \PGXzb, \PGSmb, \PGOmb, and corresponds to all \Hb that can give rise to a \PJGy meson in at least one decay mode, based on Ref.~\cite{PDG2022}, except for \PBpc decays, as discussed earlier.
Five excited $\ccbar$ resonances, \PGyP{2S}, \PGyP{3770}, \PGccDzP{1P}, \PGccDoP{1P}, and \PGccDtP{1P} are forced to decay in modes with a \PJGy meson, if they come from an \Hb. The relative proportions between different decays are preserved.
The majority (98\%) of $\Hb\to\PJGy(\to\PGmp\PGmm) + \PGmp$ backgrounds arise from the accidental pairing of a \PJGy and a \PGmp candidate from different decay chains, which can therefore populate the $m(3\PGm)>m_{\PBpc}$ CR, otherwise kinematically forbidden for both \PBcJpsitau and \PBcJpsimu decays. This signal-depleted, background-enriched region is used in the fit to determine the yield of the $\Hb$ background, without relying on the precise knowledge of the $\bbbar$ cross section.

For both MC samples, additional proton-proton collisions within the same or adjacent bunch crossings (pileup), drawn from a large minimum bias sample produced with \PYTHIA, are superimposed on the simulated $\bbbar$ event. Final-state photon radiation is generated using the \PHOTOS 3.61 toolkit~\cite{Barberio:1993qi}. The \GEANTfour package~\cite{Agostinelli:2002hh} is used to model the interaction of the generated particles with the CMS detector, and the samples are processed through the standard CMS reconstruction software.

Despite the excellent muon identification performance of CMS, the largest background source corresponds to events where the third muon comes from charged pion or kaon decays. 
This background component, referred to as misID background, is estimated \textit{in-situ} from CRs in data.
This background is reduced primarily through the optimization of the identification and isolation requirements imposed on the third muon. 
Since both signals have low multiplicity, the third muon is expected to be more isolated from additional nearby particles in signal events than in \Hb decays with several hadrons in the final state. 
The standard CMS relative muon isolation ($I^{\text{rel}}$)~\cite{CMS:2018rym}, defined as the sum of the transverse momenta of all the hadrons coming from the PV in a cone of $\Delta R = \sqrt{\smash[b]{\Delta\eta\,^2 + \Delta\phi\,^2}}=0.3$, and a multivariate muon identification algorithm~\cite{CMS:2022mgd} (ID) are used.

The shape and yield of the misID background contribution in the signal region (SR), where the third muon satisfies both the isolation ($I^{\text{rel}} < 0.2$) and identification (ID $= 1$) criteria, are predicted through a data driven method based on three background-enriched CRs: $I^{\text{rel}} < 0.2$ and ID $\neq 1$; $I^{\text{rel}} > 0.2$ and ID $= 1$; $I^{\text{rel}} > 0.2$ and ID $\neq 1$.
First, the probability for a misID hadron to satisfy $I^{\text{rel}} < 0.2$ is measured using events with ID $\neq 1$. This multiple-dimension distribution is fit with a neural network (NN) binary classifier to accurately model the dependence on several kinematical and topological observables and ensure its applicability to a different phase space.
The NN uses ten inputs, comprising the muon \pt and $\abs{\eta}$, secondary vertex displacement, the kinematical observables of the three-muon candidate, and the two classes are defined by $I^{\text{rel}} < 0.2$ or $>0.2$.
The NN output is then interpreted as a probability and used as a per-event weight to be applied to events with ID $= 1$ and $I^{\text{rel}} > 0.2$. In this region, both data and simulated samples with a genuine third muon are weighted, and the histogram of the latter is subtracted from the former to obtain the misID-background prediction in the SR. The subtraction is implemented in the fit model for each bin in the histograms, such that variations in the MC distributions are propagated to the misID background.
The method is validated in a background-dominated CR ($I^{\text{rel}} > 1.5$) by comparing the predicted and actual misID-background distributions for several kinematical observables. 

An additional combinatorial background originates from three-muon events where two unrelated, opposite-charge muons with  $m_{\Pgm\PGm}$ close to $m_{\PJGy}$ are selected.
The distributions for this background are modeled from a low  $\mMuMu$ data CR, away from the \PJGy peak.
As $q^2$ is kinematically constrained by $\mMuMu$, combinatorial dimuon events in the CR present a $q^2$ distribution different to that of those in the SR. Therefore, to predict the $q^2$ shape of this background in the SR, the four-momenta of dimuon events in the CR are scaled by the ratio of the known $m_{\PJGy}$ to the average $\mMuMu$ in the CR. The corresponding yield is derived from integrating the background function, obtained from a fit to the $\mMuMu$ distribution, in a narrow range around $m_{\PJGy}$.
This method is validated in data by comparing the observed $q^2$ distribution of the dimuon combinatorial background in the upper sideband to the prediction obtained from events in the lower sideband.

\begin{figure}[htbp]
\centering
\includegraphics[width=\prlWidth]{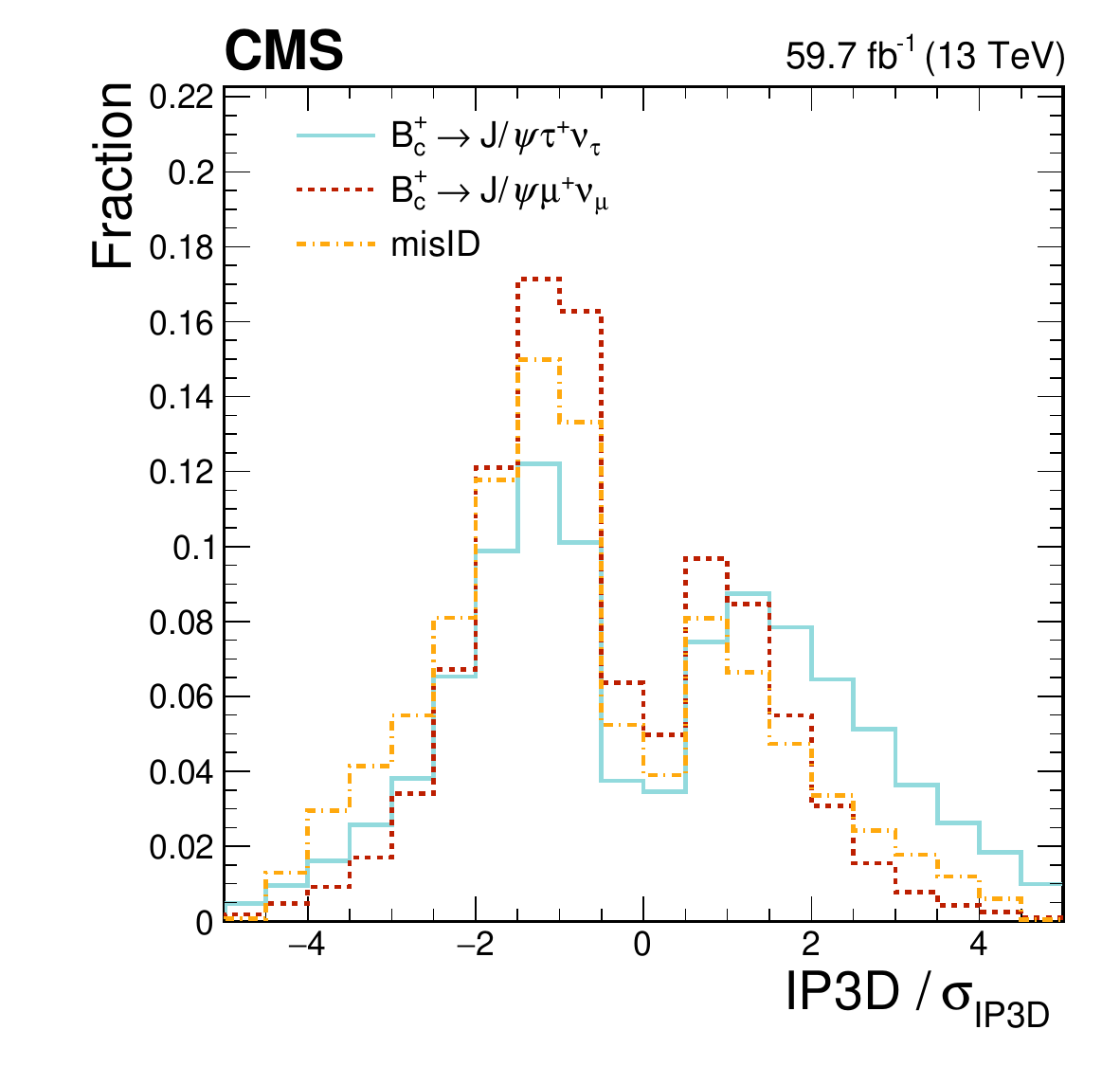}
\caption{Normalized distributions of the $\mathrm{IP3D}/\sigma_{\mathrm{IP3D}}$ observable for the \PBcJpsimu and \PBcJpsitau signal channels and the misID background. Events are selected in the region $m(3\PGm)<m_{\PBpc}$. The first and last bins contain underflow and overflow events, respectively.} \label{fig:Fit_result1}
\end{figure}

Before performing the final fit to data, events are classified in seven pairs of nonoverlapping categories.
In each pair, categories differ only by the third muon passing or failing the isolation requirement. This split is instrumental to the misID background in-situ evaluation: the fail-isolation category corresponds to the CR where misID probability weights are applied to predict the misID background in the pass-isolation SR.
The $L_{xy}/\sigma_{L_{xy}}$ or the $q^2$ observable is used in the final fit for categories designed to constrain the misID and \Hb backgrounds, or isolate the signals, respectively, as each provides the optimal separation between the relevant processes. 
The four categories enriched in \PBcJpsitau events leverage the $\mathrm{IP3D}/\sigma_{\mathrm{IP3D}}$ variable to enhance the $\Pgt$ lepton signal sensitivity, as the muon from the $\Pgt$ lepton decay is displaced with respect to the \PBpc vertex, because of the measurable $\Pgt$ lepton lifetime, as shown in Fig.~\ref{fig:Fit_result1}.
Further details are summarized in Table~\ref{tab:summary_categories}.

A binned maximum-likelihood fit is performed simultaneously to all categories, using the CMS statistical analysis tool \textsc{Combine}~\cite{CMS:2024onh}, to extract \rjpsi directly.
Since the \PBpc meson production cross section cancels in the \rjpsi ratio, a single scale parameter, which is left floating in the fit, is assigned to all \PBpc-initiated processes.
Because of the large $\PBcJpsimu$ yield, this parameter is strongly constrained in the fit and its impact is negligible.
The $\Hb$ background normalization is free to vary independently from that of the \PBpc signal and is constrained in the $m(3\PGm)>m_{\PBpc}$ category.
The combinatorial background normalization can be varied in the fit within the uncertainty obtained from the dimuon fit described earlier.

\begin{table}[htbp]
\centering
\topcaption{Summary of the category pair definitions and respective observables used in the fit. Each pair is also divided by passing or failing the isolation criterion. The first four category pairs are enriched in \PBcJpsitau, the following two enriched in \PBcJpsimu, and the last one enriched in misID and $\Hb$ background events.}
\begin{scotch}{lccc}
\multicolumn{3}{c}{Category pair definitions}                                                            & \multirow{2}{*}{Fit observable}           \\ 
$m(3\mu)$                      & $q^2$                         & $\mathrm{IP3D}/\sigma_{\mathrm{IP3D}}$  &                                           \\
\hline          
\multirow{4}{*}{${<}m_{\PBpc}$} & \multirow{4}{*}{$>$5.5$\GeV^2$} & $~< -2$                                 & \multirow{4}{*}{$q^2$}                    \\ 
                               &                               & ${-}2$ -- $0$                           &                                           \\ 
                               &                               & $0$ -- $2$                              &                                           \\ 
                               &                               & $>$2                                   &                                           \\ [\cmsTabSkip]
\multirow{2}{*}{${<}m_{\PBpc}$} & \multirow{2}{*}{$<$4.5$\GeV^2$} & $<$0                                   & \multirow{2}{*}{$L_{xy}/\sigma_{L_{xy}}$} \\
                               &                               & $>$0                                   &                                           \\ [\cmsTabSkip]
${>}m_{\PBpc}$                   & \NA                           & \NA                                     & $L_{xy}/\sigma_{L_{xy}}$                  \\ 
\end{scotch}\label{tab:summary_categories}
\end{table}

Several systematic uncertainties, which can affect both yields and shapes of the different components, are incorporated into the fit as nuisance parameters and simultaneously fitted through the likelihood profiling method~\cite{Khachatryan:2014jba}. The most relevant uncertainties pertain to the \PBpc form factors, the misID background estimate, the finite size of the MC samples, and the mismodeling of $\mathrm{IP3D}/\sigma_{\mathrm{IP3D}}$ and $L_{xy}/\sigma_{L_{xy}}$ in simulated events. 

Theoretical uncertainties in the form factors are the single dominant source of uncertainty in \rjpsi because of their effect on the $q^2$ observable.
They consist of ten independent shape variations simultaneously affecting the \PBcJpsitau and \PBcJpsimu processes, corresponding to the principal components of the correlation matrix of the form factor coefficients presented in Ref.~\cite{PhysRevD.100.094503}.
 
Uncertainties in the dominant misID background significantly affect the measurement accuracy. The misID background estimate is subject to the limited number of events in the isolation-failing CRs, the statistical uncertainty of the NN training data set and its contamination from processes with genuine muons, and the finite number of events in the validation region. Furthermore, the choice of the nominal method over possible alternatives is considered as a source of uncertainty.
These uncertainty sources are accounted for as follows.
An overall 13\% normalization uncertainty, derived from residual nonclosure in the validation test in the $I^{\text{rel}} > 1.5$ control region, is applied to all categories. The difference between the nominal and an alternative estimate (based on the probability for a misID muon to satisfy the identification requirement instead of isolation) and the effect of varying the MC normalization in the region where the misID probability is measured are included as two additional shape uncertainties. Lastly, misID statistical shape uncertainties, which can vary for each bin independently, are added to account for the impact of the limited-size sample used for the misID background estimate validation.

Uncertainties due to the limited number of events in the simulated samples are implemented as bin-by-bin shape variations. 
An uncertainty of 5\% in each of the topological quantities $\mathrm{IP3D}/\sigma_{\mathrm{IP3D}}$ and $L_{xy}/\sigma_{L_{xy}}$, estimated from discrepancies between data and MC in CRs, is also implemented as a shape uncertainty and applied to all MC-based distributions.
The former results in an event migration between the four categories with higher signal sensitivity, thus explaining its sizable impact.

Uncertainties in trigger, muon ID, and isolation efficiencies (6.6\%) are also considered in the fit. Other considered systematic uncertainties are found to be negligible.

From the maximum likelihood fit to the data, we measure 
\begin{equation*}
\rjpsi = 0.17^{+ 0.18}_{- 0.17}\stat^{+ 0.21}_{- 0.22}\syst^{+ 0.19}_{- 0.18}\thy,
\end{equation*}
where the statistical, experimental systematic, and form factor uncertainties are reported separately. A further breakdown of the systematic uncertainties is given in Table~\ref{fig:Tab_syst.}.

\begin{table}[htbp]
\centering
\topcaption{Leading systematic uncertainties for the measurement of \rjpsi. The second column reports the uncertainty type: shape (S) or normalization (N). The last column shows the resulting uncertainty in the \rjpsi measurement.}
\begin{scotch}{lccc}
  Contribution                          & Type           & Unc. $(10^{-2})$ \\
 \hline                         
  Form factor (theory)                  & S              & $19$          \\ [\cmsTabSkip]
  misID statistical                     & S (bin-by-bin) & $13$          \\
  misID systematic                                & N, S           & $8$,  $0.7$   \\ [\cmsTabSkip]
  Finite MC size                        & S (bin-by-bin) & $9$     \\[\cmsTabSkip]      
  Topological                           & S              & $9$     \\[\cmsTabSkip]
  Efficiencies                          & N              & $6$     \\
 \hline
  \multicolumn{2}{l}{Total systematic uncertainty}       & $28$    \\
\end{scotch}\label{fig:Tab_syst.}
\end{table}

\begin{figure}[htbp!]
\centering
\includegraphics[width=0.49\textwidth]{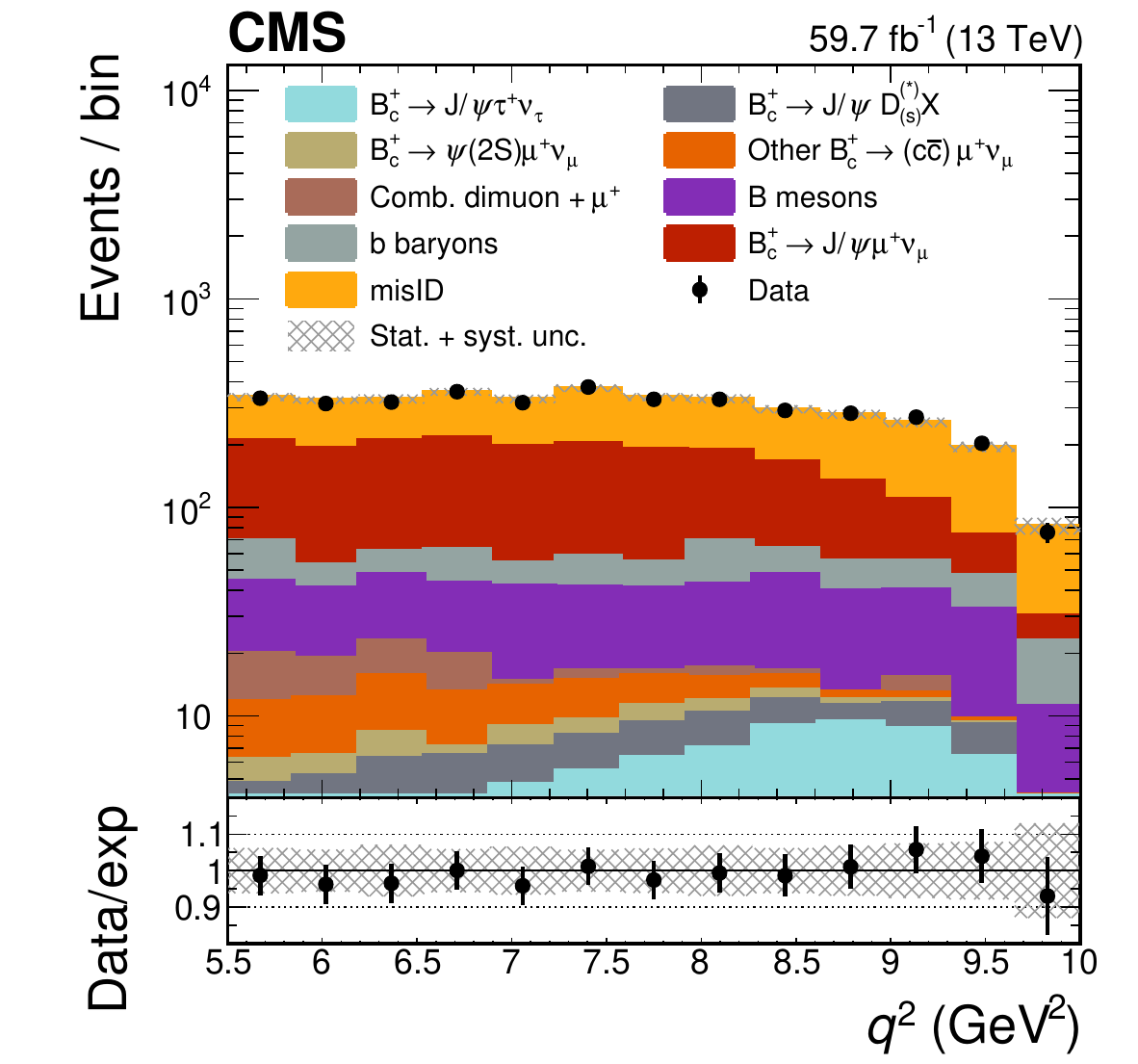}
\includegraphics[width=0.49\textwidth]{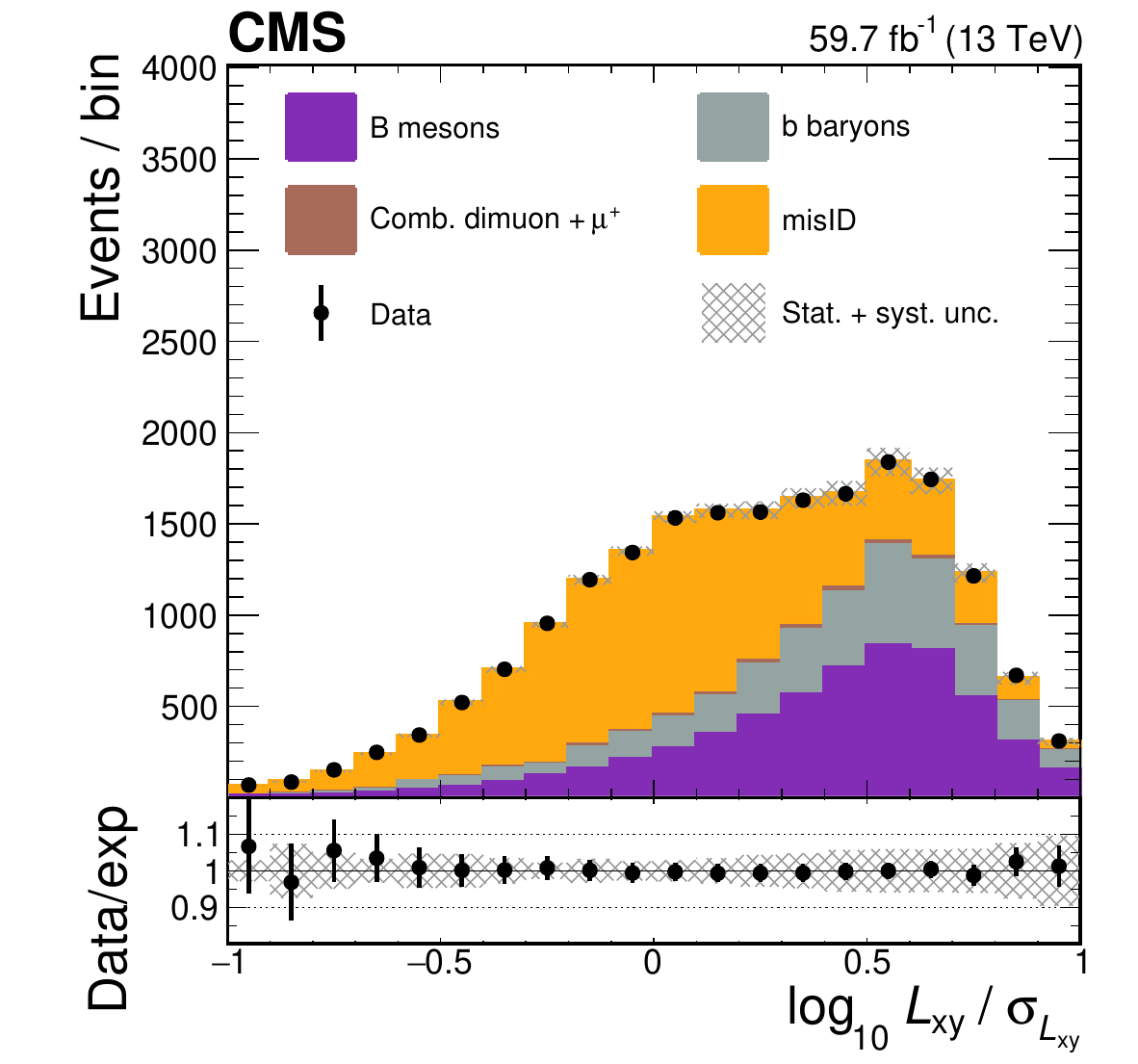}
\caption{Distributions of the $q^2$ observable in the signal-enriched category, defined by $m(3\PGm)<m_{\PBpc}$ in the bin of $q^2 > 5.5\GeV^2$ and $\mathrm{IP3D}/\sigma_{\mathrm{IP3D}}>2$ (\cmsLeft) and of the $L_{xy}/\sigma_{L_{xy}}$ observable in the category defined by $m(3\PGm)>m_{\PBpc}$ (\cmsRight). In each panel, data is compared to the best fit results.
The ratio between the data and the expected sum of signal and background contributions is shown in the lower panel of each observable. The postfit total uncertainty in the expectation is represented by the hashed band.}
\label{fig:Fit_result2}
\end{figure}

The observed yields of the \PBcJpsitau and \PBcJpsimu processes are $430\pm 570$ and $43100 \pm 4900$ events, respectively.
The distributions where the nuisance parameters, \rjpsi and their uncertainties are fixed at their best fit value (postfit) are presented in Fig.~\ref{fig:Fit_result2}, for the most signal-sensitive category (\cmsLeft) and for one background-enriched auxiliary category (\cmsRight). 

In summary, using data collected by the CMS experiment in 2018 at a center-of-mass energy of 13\TeV, corresponding to 59.7\fbinv of integrated luminosity, we measured the ratio of two branching fractions
\begin{equation*}
\rjpsi  = \frac{\mathcal{B}(\PBcJpsitau)}{\mathcal{B}(\PBcJpsimu)} = 0.17 \pm 0.33.
\end{equation*}
The result agrees with the standard model value $0.2582 \pm 0.0038$~\cite{Harrison:2020nrv} and with the LHCb measurement $0.71\pm0.17\stat\pm0.18\syst$~\cite{LHCb:2017vlu}.
This measurement is the first contribution from a general purpose experiment to the lepton flavor universality tests in \PBpc meson decays, presently only accessible at hadron colliders.

\begin{acknowledgments}
  We congratulate our colleagues in the CERN accelerator departments for the excellent performance of the LHC and thank the technical and administrative staffs at CERN and at other CMS institutes for their contributions to the success of the CMS effort. In addition, we gratefully acknowledge the computing centers and personnel of the Worldwide LHC Computing Grid and other centers for delivering so effectively the computing infrastructure essential to our analyses. Finally, we acknowledge the enduring support for the construction and operation of the LHC, the CMS detector, and the supporting computing infrastructure provided by the following funding agencies: SC (Armenia), BMBWF and FWF (Austria); FNRS and FWO (Belgium); CNPq, CAPES, FAPERJ, FAPERGS, and FAPESP (Brazil); MES and BNSF (Bulgaria); CERN; CAS, MoST, and NSFC (China); MINCIENCIAS (Colombia); MSES and CSF (Croatia); RIF (Cyprus); SENESCYT (Ecuador); ERC PRG, RVTT3 and MoER TK202 (Estonia); Academy of Finland, MEC, and HIP (Finland); CEA and CNRS/IN2P3 (France); SRNSF (Georgia); BMBF, DFG, and HGF (Germany); GSRI (Greece); NKFIH (Hungary); DAE and DST (India); IPM (Iran); SFI (Ireland); INFN (Italy); MSIP and NRF (Republic of Korea); MES (Latvia); LMTLT (Lithuania); MOE and UM (Malaysia); BUAP, CINVESTAV, CONACYT, LNS, SEP, and UASLP-FAI (Mexico); MOS (Montenegro); MBIE (New Zealand); PAEC (Pakistan); MES and NSC (Poland); FCT (Portugal);  MESTD (Serbia); MCIN/AEI and PCTI (Spain); MOSTR (Sri Lanka); Swiss Funding Agencies (Switzerland); MST (Taipei); MHESI and NSTDA (Thailand); TUBITAK and TENMAK (Turkey); NASU (Ukraine); STFC (United Kingdom); DOE and NSF (USA).
\end{acknowledgments}

\bibliography{auto_generated}
\cleardoublepage \appendix\section{The CMS Collaboration \label{app:collab}}\begin{sloppypar}\hyphenpenalty=5000\widowpenalty=500\clubpenalty=5000
\cmsinstitute{Yerevan Physics Institute, Yerevan, Armenia}
{\tolerance=6000
A.~Hayrapetyan, A.~Tumasyan\cmsAuthorMark{1}\cmsorcid{0009-0000-0684-6742}
\par}
\cmsinstitute{Institut f\"{u}r Hochenergiephysik, Vienna, Austria}
{\tolerance=6000
W.~Adam\cmsorcid{0000-0001-9099-4341}, J.W.~Andrejkovic, T.~Bergauer\cmsorcid{0000-0002-5786-0293}, S.~Chatterjee\cmsorcid{0000-0003-2660-0349}, K.~Damanakis\cmsorcid{0000-0001-5389-2872}, M.~Dragicevic\cmsorcid{0000-0003-1967-6783}, P.S.~Hussain\cmsorcid{0000-0002-4825-5278}, M.~Jeitler\cmsAuthorMark{2}\cmsorcid{0000-0002-5141-9560}, N.~Krammer\cmsorcid{0000-0002-0548-0985}, A.~Li\cmsorcid{0000-0002-4547-116X}, D.~Liko\cmsorcid{0000-0002-3380-473X}, I.~Mikulec\cmsorcid{0000-0003-0385-2746}, J.~Schieck\cmsAuthorMark{2}\cmsorcid{0000-0002-1058-8093}, R.~Sch\"{o}fbeck\cmsorcid{0000-0002-2332-8784}, D.~Schwarz\cmsorcid{0000-0002-3821-7331}, M.~Sonawane\cmsorcid{0000-0003-0510-7010}, S.~Templ\cmsorcid{0000-0003-3137-5692}, W.~Waltenberger\cmsorcid{0000-0002-6215-7228}, C.-E.~Wulz\cmsAuthorMark{2}\cmsorcid{0000-0001-9226-5812}
% ...
\par}
\vskip\cmsinstskip
\dag:~Deceased\\
$^{1}$Also at Yerevan State University, Yerevan, Armenia\\
$^{2}$Also at TU Wien, Vienna, Austria\\
% ...
\end{sloppypar}
\end{document}